\documentclass[12pt,showpacs]{revtex4}
\usepackage[english]{babel}
\usepackage{graphicx}
\usepackage{epsfig}

\begin{document}
\title{The polarizability of the pion:\\
no conflict between dispersion theory and chiral perturbation theory} \normalsize
\author{B. Pasquini}
\affiliation{
Dipartimento di Fisica Nucleare e Teorica, Universit\`{a} degli Studi di Pavia, and\\
Istituto Nazionale di Fisica Nucleare, Sezione di Pavia, I-27100 Pavia, Italy}
\author{D. Drechsel and S. Scherer}
\affiliation{ Institut f\"ur Kernphysik, Johannes Gutenberg-Universit\"at,
D-55099 Mainz, Germany}
\begin{abstract}
Recent attempts to determine the pion polarizability by dispersion relations
yield values that disagree with the predictions of chiral perturbation theory.
These dispersion relations are based on specific forms for the absorptive part
of the Compton amplitudes. The analytic properties of these forms are examined, and
the strong enhancement of intermediate-meson contributions is shown to be
connected with spurious singularities. If the basic requirements of dispersion relations
are taken into account, the results of dispersion theory and effective field theory are
not inconsistent.
\end{abstract}
\date{\today}
\maketitle
\section{Introduction}
\label{sec:introduction}
The polarizabilities of a composite system such as the pion are
elementary structure constants, just as its size and shape.
They can be studied by applying electromagnetic fields to the
system. The physical content of the polarizability can be
visualized best by effective multipole interactions for the
coupling of the electric $(\vec{E})$ and magnetic $(\vec{H})$
fields of a photon with the internal structure of the pion.
This structure can be accessed experimentally by the Compton
scattering process $\gamma + \pi \to \gamma + \pi$ or the
crossed-channel reaction $\gamma + \gamma \to \pi + \pi$. When
expanding the Compton scattering amplitude in the energy of the
photon, the zeroth- and first-order terms follow from a low-energy theorem
and can be expressed solely in terms of the
charge and mass of the pion. The second-order terms in the
photon energy describe the response of the pion's internal
structure to an external electric or magnetic dipole field,
they are proportional to the electric ($\alpha)$ and magnetic
($\beta$) dipole polarizabilities, respectively. Expanding the
Compton  amplitudes to higher orders in the energy, one obtains
higher-order polarizabilities, e.g., the quadrupole polarizabilities at fourth order.\\

From the theoretical side there is an extraordinary interest in
a precise determination of the pion polarizabilities. Within
the framework of the partially conserved axial-vector current (PCAC)
hypothesis and current algebra the electromagnetic
polarizabilities of the charged pion are related to the
radiative decay $\pi^+\to e^+\nu_e\gamma$~\cite{Terentev:1972ix}.
The result obtained using chiral perturbation theory (ChPT) at
leading non-trivial order (${\cal O}(p^4)$)~\cite{Bijnens:1987dc}
is equivalent to the original PCAC
result, $\alpha_{\pi^+}=-\beta_{\pi^+}\sim \bar l_\Delta$,
where $\bar l_\Delta\equiv(\bar l_6-\bar l_5)$ is a linear
combination of scale-independent parameters of the Gasser and
Leutwyler Lagrangian~\cite{Gasser:1983yg}. At ${\cal O}(p^4)$
this difference is related to the ratio $\gamma=F_A/F_V$ of the
pion axial-vector form factor $F_A$ and the vector form factor
$F_V$ of radiative pion beta decay~\cite{Gasser:1983yg},
$\gamma={\bar{l}}_\Delta/6$. Once this ratio is known, chiral symmetry
makes an {\em absolute} prediction for the polarizabilities.
Using the most recent determination $\gamma=0.443\pm 0.015$ by
the PIBETA Collaboration~\cite{Frlez:2003pe} (assuming
$F_V=0.0259$ obtained from the conserved vector current
hypothesis) results in the ${\cal O}(p^4)$ prediction
$\alpha_{\pi^+}=2.64\pm 0.09$ in units of $10^{-4}\,
\mbox{fm}^3$, where the estimate of the error is only the one
due to the error of $\gamma$ and does not include effects from
higher orders in the quark mass expansion.\\

Corrections to the leading-order PCAC result have been
calculated at ${\cal O}(p^6)$ in chiral perturbation theory and
turn out to be rather small~\cite{Burgi:1996qi,Gasser:2006qa}.
Contrary to the situation of the nucleon, no ``matter fields''
with their own mass scale are present, and therefore the
calculations can be performed in the original formulation of
ChPT~\cite{Gasser:1983yg}. This makes the following predictions
for the polarizabilities a very significant test of this
theory~\cite{Gasser:2006qa}:
\begin{eqnarray}
\alpha_{\pi^+} + \beta_{\pi^+} &=& 0.16 \,,\label{eq:1.1}\\
\alpha_{\pi^+} - \beta_{\pi^+} &=& 5.7 \pm 1.0\,.\label{eq:1.2}
\end{eqnarray}
The error for $\alpha_{\pi^+} + \beta_{\pi^+}$ is
of the order 0.1, mostly from the dependence on the scale at which the
${\cal O}(p^6)$ low-energy coupling constants are estimated by resonance saturation.
The forward polarizability could also obtain relatively large contributions
at ${\cal O}(p^8)$. On the other hand, there is as yet no indication
of large higher-order effects for the backward polarizability
$\alpha_{\pi^+} - \beta_{\pi^+}$. For further information on low-energy
$\pi \gamma$ reactions we refer to the recent review by Kaiser and
Friedrich~\cite{Kaiser:2008ss}.\\

The pion polarizability  has been studied in lattice QCD~\cite{Wilcox:1996vx}.
The valence-quark contribution to the
electric polarizability was shown to yield only the small value $\alpha_{\pi^+}\approx -0.17$,
one order of magnitude smaller than the value predicted by ChPT. It is, of course, not surprising
that sea quarks and their correlations must play an important role, most likely
configurations with the quantum numbers of the pion. In a recent contribution,
Hu~{\emph {et al.}} also conclude that polarizabilities are difficult to predict in
lattice QCD because of (partial) quenching and volume effects~\cite{Hu:2007ts}. However, these authors
point out that forthcoming lattice QCD results can be used as a diagnostic for ChPT. \\

The results of ChPT are in sharp contrast with the predictions of
Fil'kov and Kashevarov~\cite{Fil'kov:1982cx,Fil'kov:1998np,
Fil'kov:2005ss,Fil'kov:2005yf} who obtain
\begin{equation}
11.1 \leq \alpha_{\pi^+} - \beta_{\pi^+} \leq 15.6\label{eq:1.2FK}
\end{equation}
in recent work~\cite{Fil'kov:2005ss} based on dispersion relations (DRs).
The dispersion integrals are saturated by various meson contributions in the
$s$ and $t$ channel. The free parameters are essentially fixed by the
known masses, total widths, and partial decay widths of these mesons at
resonance. However, the extrapolation to energies below and above the resonance
is performed with specific resonance shapes whose analytic properties
leave room for a considerable model dependence.\\

The very small value predicted by Eq.~(\ref{eq:1.1}), that is
Baldin's sum rule applied to the pion, makes a measurement of
this observable close to impossible. The experiments are
therefore analyzed with the constraint $\alpha_{\pi^+} =
-\beta_{\pi^+}$. Unfortunately, the experimental situation is
rather contradictory, see Refs.~\cite{Ahrens:2004mg,Gasser:2006qa}
for recent reviews of the
data and further references to the experiments. There exist
basically three different methods to measure $\alpha_{\pi^+}$:
(I) the reactions $e^+e^- \rightarrow \gamma \gamma \rightarrow
\pi^+\pi^-$, (II) the Primakov effect of scattering a
relativistic pion in the Coulomb field of a heavy nucleus, and
(III) the radiative pion photoproduction, $p(\gamma, \gamma'
\pi^+ n)$, which contains Compton scattering on an off-shell pion
as a subprocess. The latter reaction was recently investigated
at the Mainz Microtron MAMI with the result~\cite{Ahrens:2004mg}
\begin{equation}
\alpha_{\pi^+} - \beta_{\pi^+} = 11.6 \pm 1.5_{\text{stat}} \pm
3.0_{\text{syst}} \pm 0.5_{\text{mod}}\,, \label{eq:1.3}
\end{equation}
which is at variance with the prediction of Gasser {\emph {et
al.}}~\cite{Gasser:2006qa} by two standard deviations. In view of the
theoretical uncertainties from the fact that the photon is scattered by an
off-shell pion, the deviation from theory is an open problem. In particular, we
point out that the model error in Eq.~(\ref{eq:1.3}) is estimated by comparing
the analysis with 2 specific models. This does not exclude that a wider range
of models will lead to larger model errors. Because the pion polarizability is extremely
important for our understanding of QCD in the confinement region, it is
prerequisite to check the given arguments by a full-fledged ChPT calculation of
the reaction $p(\gamma,\gamma' \pi^+n)$.\\

The second method to determine the polarizability, the Primakov effect, has
been studied at Serpukhov with the result~\cite{Antipov:1982kz}
\begin{equation}
\alpha_{\pi^+} - \beta_{\pi^+} = 13.6 \pm 2.8_{\text{stat}} \pm
2.4_{\text{syst}}\,, \label{eq:_1.4}
\end{equation}
in agreement with the value from MAMI. Recently, also the COMPASS Collaboration
at CERN has investigated this reaction, and the data analysis is
underway~\cite{Colantoni:2005ku,Abbon:2007pq}.\\

Unfortunately, the third method based on the
reactions $e^+e^- \rightarrow \gamma \gamma \rightarrow \pi^+\pi^-$, has led to
even more contradictory results in the range $4.4 \leq \alpha_{\pi^+}\leq
52.6$, as listed in the work of Gasser {\emph {et al.}}~\cite{Gasser:2006qa}. Therefore,
one has to wait for an improved analysis of the data before
final conclusions can be drawn.  At the same time new and independent
experimental effort is invaluable, such as the planned experiment at Jefferson Lab
after the 12~GeV upgrade.\\

In this work we address the conflicting results obtained by ChPT and DRs.
Section~\ref{sec:kinematics} gives a brief introduction to the kinematics and
scattering amplitudes relevant for these studies. In Sec.~\ref{sec:models} we
summarize the elements of previous calculations in the framework of dispersion
relations. Moreover, the approximations involved are critically investigated
within several simple but pertinent approximations. Our results for dispersion
relations in the $t$ channel are presented in Sec.~\ref{sec:t-channel}.
Finally, we give our conclusions in Sec.~\ref{sec:conclusion}.
%
%
\section{Kinematics and scattering amplitudes}
\label{sec:kinematics}

\begin{figure}[ht]
\begin{center}
\epsfig{file=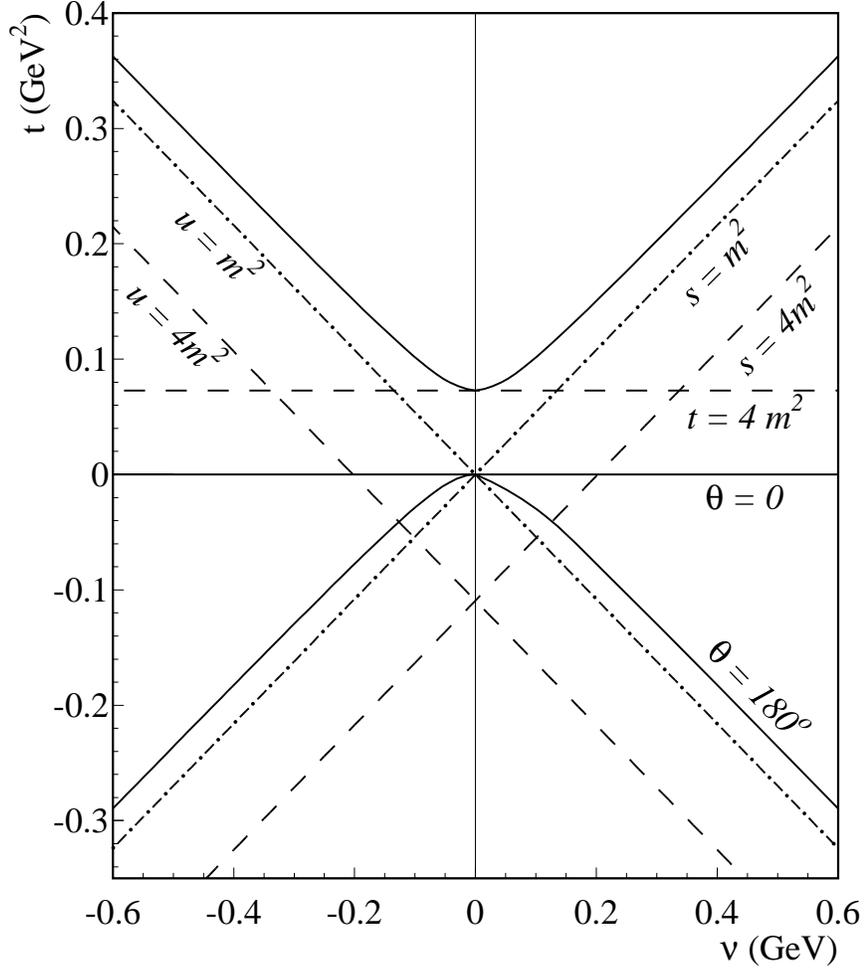,width=12 cm}
\end{center}
\caption{The Mandelstam plane for Compton scattering off the pion. Solid lines:
boundaries of the physical  $s$, $u$, and $t$ channels, in particular forward
($\theta = 0$) and backward ($\theta = 180^{\circ}$) scattering in the $s$ channel;
dashed lines: inelastic thresholds in the three channels; dash-dotted lines:
possible paths for DRs at $s=m^2$ and $u=m^2$. Note: forward DRs are integrated along
the $\nu$ axis, i.e., at $t=0$.}
\label{fig:mandelstam}
\end{figure}

Let us consider the kinematics of Compton scattering, the reaction $\gamma(k) +
\pi(p) \rightarrow \gamma(k') + \pi(p')$, where  the variables in brackets
denote the 4-momenta of the participating particles. The familiar Mandelstam
variables are
\begin{equation}
s=(k+p)^2\ ,\ \ t=(k-k')^2\ ,\ \ u=(k-p')^2\ , \label{eq2.1}
\end{equation}
which are constrained by $s+t+u=2m^2$, where $m$ is the pion mass. The
crossing-symmetric variable $\nu$ is defined by
\begin{equation}
\nu=\frac{s-u}{4m}\, . \label{eq2.2}
\end{equation}
The two Lorentz-invariant variables $\nu$ and $t$ span the Mandelstam plane shown in
Fig.~\ref{fig:mandelstam}. They are related to the initial ($E_\gamma$) and final
($E'_\gamma$) photon lab energies and to the lab scattering angle $\theta$  by
\begin{eqnarray}
\nu&=&E_\gamma+\frac{t}{4m}=\frac{1}{2}(E_\gamma+E'_\gamma),\nonumber\\
t&=&-4E_\gamma \, E'_\gamma \, \sin^2 (\theta /2)= -2m(E_\gamma-E'_\gamma).
\label{eq2.3}
\end{eqnarray}
The scattering matrix of Compton scattering on the pion, $T$, can be expressed
by 2 independent amplitudes $A_i(\nu, t)$, $i=1,2$. These structure functions
depend on $\nu$ and $t$, they are free of kinematic singularities and
constraints, and because of the crossing symmetry they satisfy the relation
$A_i(\nu, t)=A_i(-\nu, t)$.  We further note that the functions $A_i$ are real
in the interior of a triangle formed by the dashed lines $s=t=u=4 m^2$ in
Fig.~\ref{fig:mandelstam}. In the following we use these amplitudes to set up DRs.
The amplitudes $A_i$ are related to the amplitudes $T_i$ of
Prange~\cite{Prange:1958dd} as follows:
\begin{eqnarray}
T_1 &=& {\textstyle \frac {1}{2}}(t A_1+\eta A_2)\,, \nonumber\\
T_2 &=& {\textstyle \frac {1}{2}}(t A_1-\eta A_2)\,,\label{eq2.4}
\end{eqnarray}
with $\eta= 4\nu^2+t-t^2/(4m^2)=(m^4-su)/m^2=((s-m^2)^2+st)/m^2$. In terms of
these amplitudes, the $T$ matrix takes the form
\begin{equation}
T=\frac {\varepsilon' \cdot P'\,\varepsilon \cdot P'}{P' \cdot P'}T_1 + \frac
{\varepsilon' \cdot N\,\varepsilon \cdot N}{N \cdot N}T_2 \,, \label{eq2.5}
\end{equation}
where $\varepsilon$ and $\varepsilon'$ are the photon polarization four-vectors in
the initial and final states, respectively. Furthermore we
have defined the following 4-vectors
\begin{eqnarray}
K\; &=& {\textstyle \frac {1}{2}}(k'+k)\,, \; P = {\textstyle \frac
{1}{2}}(p'+p)\,, \; Q = {\textstyle \frac {1}{2}}(k'-k)\,, \nonumber\\
P' &=& P-\frac {P \cdot K}{K \cdot K} K \,, \quad N^{\mu} = \epsilon^{\mu
\alpha \beta \gamma}P'_{\alpha}Q_{\beta}K_{\gamma}\,, \label{eq2.6}
\end{eqnarray}
with $\epsilon_{0123}=+1$. The differential cross section for Compton
scattering is constructed from the $T$ matrix by
\begin{eqnarray}\label{cs-generic}
\left(\frac{d\sigma}{d\Omega}\right)_{\gamma\pi}= \Phi^2\left|T\right|^2,
\qquad \mbox{with} \qquad \Phi = \cases{  \displaystyle \frac {1}{8 \pi
m}\frac{E'_\gamma}{E_\gamma}  & (lab frame) \cr \displaystyle \frac {1}{8 \pi
\sqrt{s}}  & (c.m. frame), }
\end{eqnarray}
where
\begin{eqnarray}\label{Tsquare}
|T|^2 &=&  \frac{1}{4} \Big( t^2 |A_1|^2 + \eta^2 |A_2|^2\Big).
\end{eqnarray}
We further note that the $t$-channel reaction $\gamma \gamma \rightarrow \pi \pi$
is usually described
by the amplitudes $M_{++}=-A_1/2$ and $M_{+-}=-A_2/(2m^2)$, with
indices referring to the polarization of the incident photons. These
amplitudes describe the respective cross section as follows:
\begin{eqnarray}
\left(\frac{d\sigma}{d\Omega}\right)_{\gamma\gamma} &=& \frac {1}{128 \pi^2}
\sqrt{\frac{t-4m^2}{t^3}}\bigg\{t^2 \mid M_{++} \mid^2\nonumber\\
&+&\frac {1}{16}t^2(t-4m^2)^2\sin^4{\theta^{\ast}}\mid M_{+-}
\mid^2\bigg\}\,,\label{gamma_gamma_pi_pi}
\end{eqnarray}
with $\theta^{\ast}$ the angle between the incident photon and the outgoing
pion in the c.m.~frame. The cross section for $\gamma + \gamma \rightarrow
\pi^0 + \pi^0$ is obtained by multiplying the r.h.s of
Eq.~(\ref{gamma_gamma_pi_pi}) with a factor 1/2, which accounts for two
identical particles in the final state.\\

Assuming analyticity and an appropriate high-energy behavior, the amplitudes
$A_i$ fulfill unsubtracted DRs at fixed $t$,
\begin{equation}
{\rm Re} A_i(\nu, t) \;=\; A_i^B(\nu, t) \;+\; {2 \over \pi} \; {\mathcal P}
\int_{\nu_{\mathrm {thr}}(t)}^{+ \infty} d\nu' \; \frac{\nu' \; \mathrm{Im}_s
A_i(\nu',t)} {\nu'^2 - \nu^2}\;. \label{eq:DRnu}
\end{equation}
The Born terms $A_i^B$ describe photon scattering off a point-like pion, which leads
to poles for $s=m^2$ and $u=m^2$ shown by the dash-dotted lines in Fig.~\ref{fig:mandelstam}.
In terms of the orthogonal coordinates these pole lines are given by
$\nu (t) = \pm \nu_B (t) = \pm t/4m$. Furthermore,
$\mathrm{Im}_s A_i$ are the discontinuities across the $s$-channel cut of
the Compton process and $\nu_{\text {thr}} (t)={\textstyle \frac{3}{2}m} +\nu_B (t)$
is the threshold for two-pion production in the $s$ channel. The sum of pole and contact terms
takes the form
\begin{equation}
A_1^B(\nu, t) = A_2^B(\nu, t) = \frac {e^2\,q}{(\nu -\nu_B)(\nu +\nu_B)}\,,\label{eq:Born}
\end{equation}
where $q$ is 0 for neutral and 1 for charged pions. As is obvious from Eq.~(\ref{eq:Born}),
the Born contributions to the invariant amplitudes have a pure pole structure.
The $s$-channel cut starts at the lowest
production threshold, which is given by intermediate two-pion states, i.e.,
$s_{\text {thr}} = 4 m^2$. The same cuts appear in the $t$ and $u$ channels
(see Fig.~\ref{fig:mandelstam}). For DRs at constant $t\leq 0$, the
crossing symmetry allows one to combine the $s$- and $u$-channel contributions
in the form of Eq.~(\ref{eq:DRnu}).\\

Other types of DRs evade the $u$-channel
contributions and replace them by the discontinuity in the $t$ channel. These
are DRs at fixed $u = m^2$ or at constant angle, e.g., $\theta = 180^{\circ}$.
The former DRs take the form~\cite{Fil'kov:1998np}
\begin{equation}
{\rm Re} A_i(s, t) \;=\; A_i^B(s, t) +\; \frac{1}{\pi} \; {\mathcal P}\bigg\{
\int_{t_{\text {thr}}}^{+ \infty} dt' \; \frac{\mathrm{Im}_t
A_i(t',u=m^2)}{t'-t} + \int_{s_{\text {thr}}}^{+ \infty} ds' \;
\frac{\mathrm{Im}_s A_i(s',u=m^2)}{s'-s}\bigg\} \;, \label{eq:DRu}
\end{equation}
with the constraint $s+t=m^2$. In particular, the polarizabilities are obtained
at the point $s=m^2$ and $t=0$,
\begin{equation}
\alpha + \beta = -\frac {1}{4 \pi m}\, A_2^{\text
{disp}}(s=m^2,\,t=0)\,,\quad\alpha - \beta = -\frac {1}{4 \pi m}\,
A_1^{\text {disp}}(s=m^2,\,t=0), \label{eq:alpha_beta}
\end{equation}
with $A_i^{\text {disp}}$ the dispersive, i.e., non-pole contribution to the respective amplitude. For
further convenience we introduce the dynamic polarizabilities
\begin{equation}
P^{(+)}(s,t)= -\frac {1}{4 \pi m}\,A_2^{\text {disp}}(s,t)\,, \quad
P^{(-)}(s,t)= -\frac {1}{4 \pi m}\,
A_1^{\text{disp}}(s,t)\,,\label{eq:dyn_pol}
\end{equation}
with $P^{(\pm)}(m^2,0)=\alpha \pm \beta$.\\

In order to evaluate the dispersion integral in Eq.~(\ref{eq:DRnu}),
the imaginary parts in the $s$ channel are  determined by the unitarity relation, taking
account of two-pion states and resonance contributions such as vector mesons in the
intermediate state. The $t$-channel contribution of Eq.~(\ref{eq:DRu}) is obtained
in terms of unitarized partial-wave amplitudes following the method outlined in
Refs.~\cite{Drechsel:1999rf,Drechsel:2002ar}. In the $t$ channel, the amplitudes
$A_1$ (or $M_{++}$) and $A_2$ (or $M_{+-}$) correspond to photon helicity differences of
$\Lambda_\gamma=0$ and $\Lambda_\gamma=2$, respectively.
\section{Models}
\label{sec:models}
In this section we study 3 generic resonance models with the amplitudes
\begin{eqnarray}
F_A(s)&=& \frac {1}{M^2-s-{\textstyle {\frac {1}{4}}}\Gamma_0^2-iM\Gamma_0}\,,\label{eq:FA}\\
F_B(s)&=& \frac {1}{M^2-s-{\textstyle {\frac {1}{4}}}[\Gamma(s)]^2-iM\Gamma(s)}\,,\label{eq:FB}\\
F_C(s)&=& \frac {1}{M^2-s-iM\Gamma(s)}\label{eq:FC}
\end{eqnarray}
with
\begin{equation}
\Gamma(s) = \bigg(\frac {s-s_0}{M^2-s_0}\bigg)^{\frac
{3}{2}}\;\Gamma_0. \label{eq:Gamma-s}
\end{equation}
The amplitude $F_A$ describes an ideal resonance, a fixed pole at
$s=(M-{\textstyle {\frac {i}{2}}}\;\Gamma_0)^2$. The amplitudes $F_B$
and $F_C$ have an energy-dependent width leading to a branch cut
in $s$ from $s_0=4m^2$ to $+\infty$. The onset of the cut is given by the
threshold for two-pion production, and the energy dependence of the
width in Eq.~(\ref{eq:Gamma-s}) corresponds to a $P$-wave resonance,
i.e., intermediate vector mesons like $\rho$ or $\omega$. For a
general angular momentum $L$, the width opens like
$(s-s_0)^{(2L+1)/2}$. The amplitude $F_C$ is obtained in the
small-width approximation, $\Gamma_0 \ll M$, this form is used in
the work of Fil'kov and
Kashevarov~\cite{Fil'kov:2005ss,Fil'kov:2005yf}. Furthermore, these
authors introduce an energy-dependent coupling constant for the
excitation and the decay of the intermediate vector meson,
\begin{equation}
g^2(s) = 6 \pi \sqrt{\frac{M^2}{s}} \bigg ( \frac {M}{M^2-m^2} \bigg
)^3\,\Gamma_{\gamma \pi}\,, \label{eq:g2-s}
\end{equation}
where $\Gamma_{\gamma \pi}$ is the partial width for the decay of
the vector meson to a pion-photon state. Combining the above equations
with Eq.~(\ref{eq:dyn_pol}), we find the following expressions for the
dynamic polarizabilities:
\begin{equation}
P^{(+)}_N(s)= \frac {2 m}{\pi}\,g^2(s)\,F_N(s),\quad P^{(-)}_N(s)=
-\frac {2 s}{\pi m}\,g^2(s)\,F_N(s)=-\frac{s}{m^2}\,P^{(+)}_N(s)\,.
\label{eq:models}
\end{equation}
This defines the models $A$, $B$, and $C$ discussed in the following. Due
to the square-root singularity at the origin, the coupling of
Eq.~(\ref{eq:g2-s}) leads to an (unphysical) cut from $s=-\infty$ to
$s=0$. For this reason we also introduce models with an
energy-independent coupling fixed by the value at the resonance
position, $g_0^2=g^2(M^2)$. These models are denoted by $A0$, $B0$,
and $C0$. If all the requirements to set up the DRs are fulfilled,
the direct calculation of the polarizabilities
from the real part of the Compton amplitude has to yield the same
result as obtained from the DRs with the imaginary part as input.
For a comparison with the physics, the numerical calculations are
performed for $m=M_{\pi+}=0.140$~GeV, $M=M_{\rho}=0.770$~GeV,
$\Gamma_0=\Gamma_{\rho}=0.151$~GeV, and $\Gamma_{\gamma
\pi}=\Gamma_{\rho \gamma \pi}=0.068$~MeV.
\subsection{Forward polarizability}
\begin{figure}[ht]
\begin{center}
\epsfig{file=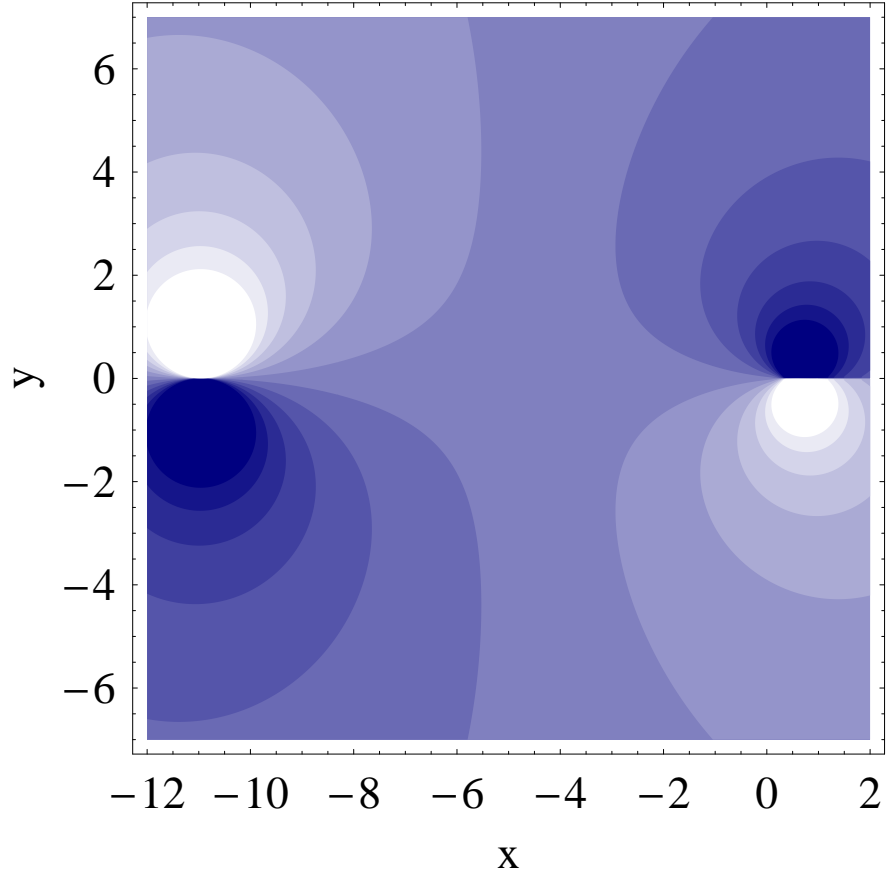,width=12 cm}
\end{center}
\caption{Contour plot of the imaginary part of the dynamical polarizability $P^{(+)}_{C0}$ in
the complex $s=x+{\text {i}}\,y$ plane, with $x$ and $y$ in units of ${\text {GeV}}^2$. The cut
due to two-pion intermediate states (right) leads from
$x \approx 0.02$, $y=0$ to $x \rightarrow \infty$, $y=0$.
The spurious pole of this model is seen on the left near $x=-11$.} \label{fig:modelC0}
\end{figure}

\begin{figure}[ht]
\begin{center}
\epsfig{file=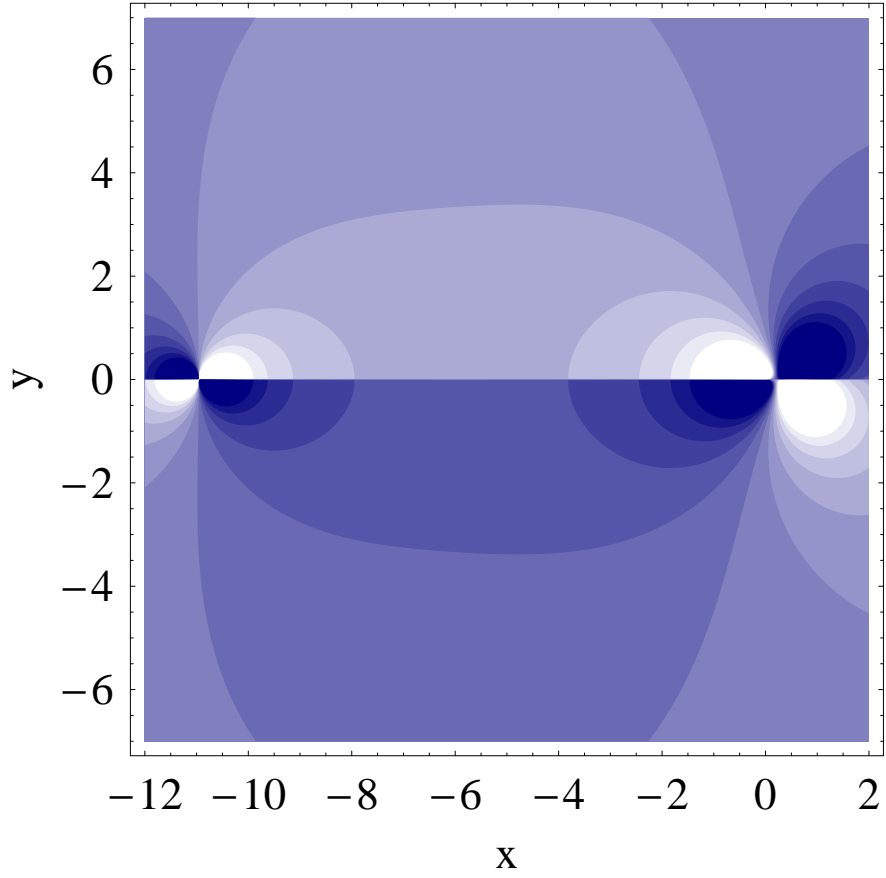,width=12 cm}
\end{center}
\caption{Contour plot of the imaginary part of the dynamical polarizability $P^{(+)}_{C}$ in
the complex $s=x+{\text {i}}\,y$ plane, with $x$ and $y$ in units of ${\text {GeV}}^2$. The cut
due to two-pion intermediate states (right) leads from
$x \approx 0.02$, $y=0$ to $x \rightarrow \infty$, $y=0$. The (unphysical) left cut runs from
$x = 0$, $y=0$ to $x \rightarrow -\infty$, $y=0$. The spurious pole of Fig.~\ref{fig:modelC0}
appears embedded on the left cut near $x=-11$.} \label{fig:modelC}
\end{figure}

Let us first address the DR for the forward
polarizability, $P^{(+)}$, which is obtained from the dispersion integral Eq.~(\ref{eq:DRnu})
evaluated at $t=0$ (forward DR). The results for the $s$-channel
contribution are listed in Table~\ref{table1}. If the $u$ channel is also
included, the full polarizability is obtained by
multiplication with a factor 2 (crossing symmetry). Because the amplitude $F_{A0}$ does
not have a cut, the dispersion integral runs over the full real
axis. In the table, the integrals have been divided into the
contributions above and below the physical branch point $s=s_0$,
labeled ``right cut'' and ``left cut'', respectively. The
(unphysical) contribution of model $A0$ below the production
threshold is clearly rather small. A look at the table shows that
the models yield quite similar contributions from the right cut.
However, the models differ substantially in their analytic
structure. In particular, they differ because of the following
ingredients:
\begin{itemize}
\item
The energy-dependent width of models $B$, $B0$, $C$, and $C$0 introduces the correct
physical cut starting at $s=4m^2$ and shifts the resonance pole to the second Riemann
sheet. At the same time, this energy dependence leads to (unphysical) singularities of
the amplitude on the first sheet: a pair of complex-conjugate poles in models $B0$ and
$B$ as well as a (spurious) pole on the negative axis at $s \approx -11~{\text {GeV}}^2$ in model $C0$,
that is, a deeply bound state of the pion-photon system. Independent of the physical
questions involved with these models, we can not simply ignore these cuts and poles.
As an example, Fig.~\ref{fig:modelC0} displays the imaginary part of the dynamic
polarizability $P_{C0}^{(+)}$. Because the static polarizability is obtained at the
small value $s=m^2$, close to the onset of the cut seen on the right side of the figure,
the effect of the distant pole on the left is suppressed (see Table~\ref{table1}).
\item
The energy-dependent coupling constant in models $A$, $B$, and $C$ leads
to an even more serious problem, a $1/\sqrt{s}$ singularity
at the origin resulting in a further cut. In order to obtain real values
for the polarizabilities, we draw this cut from $s=0$ to $-\infty$ (left cut).
Obviously, the square-root singularity is very close
to the threshold of Compton scattering, $s=m^2$, in which point the
polarizability is determined. Within the physical (right) cut, an energy
dependence of the amplitudes with $1/\sqrt{s}$ may be a reasonable approximation,
it increases the spectral function at energies below the
resonance and damps the meson production at very large energies.
However, the identification of the right-cut dispersion integral
with the full amplitude at the Compton threshold is questionable, because the model
introduces unphysical properties on the first sheet of the complex $s$ plane.
In fact the left-cut contribution is dominant and leads to a considerable
increase of the polarizabilities by a factor of $M/m$. As an example, the imaginary part
of the dynamic polarizability $P_{C}^{(+)}$ is displayed in Fig.~\ref{fig:modelC}. The right
cut in this figure has not changed much as compared to Fig.~\ref{fig:modelC0}, whereas the
additional left cut with the embedded pole structure at $s \approx -11~{\text {GeV}}^2$ has
become the dominant feature. In particular, the onset
of the spurious left cut at $s=0$ appears very close to the Compton threshold at $s=m^2$,
and the consequences are seen in Tables~\ref{table1} and \ref{table2}.
\end{itemize}
In a somewhat different language, the polarizability $P(m^2)$ can be
represented by a contour integral along a small circle about $s=m^2$
with radius $\epsilon$,
\begin{equation}
P(m^2) \;=\;\frac {1}{2 \pi i} \oint \frac {P(s')}{s'-m^2} ds'\,.
\label{eq:circle}
\end{equation}
If we blow up the contour, we expect only contributions from the
upper and lower rim of the cut from $s=s_{\text {thr}}$ to $+\infty$
and, possibly, from a full circle at infinity. Because the above
amplitudes converge better than $1/s$ for large $s$, the
contribution of the large circle converges to zero. Obviously,
unphysical poles on the physical sheet and the unphysical left
cut are obstacles for blowing up the contour, and therefore we have
to add (I) the residues of the integrand $P(s')/(s'-s)$ at the poles
and (II) the integral over the discontinuity of the imaginary part
on the left cut in order to agree with the value given by the
real part of the Compton amplitude. We hasten to add that all the
above models have unphysical properties and therefore deserve
further studies. However, it is our point to demonstrate that the
experimentally known amplitude at resonance, $s=M^2$, can not be
uniquely continued to the threshold for Compton scattering, $s=m^2$,
but that such procedure leaves room for large model errors. It is
also worth pointing out that in models $B$ and $C$ the crossing symmetry
leads to four cuts, which cover the complete real axis with the result of
complex (quasi-static) polarizabilities .\\
\subsection{Backward polarizability}
Let us now turn to the dispersion relation for the backward
polarizability $P^{(-)}$. Compared to the forward polarizability
$P^{(+)}$, the additional factor $s$ in Eq.~(\ref{eq:models}) leads to
a much slower
convergence of the integrals. This is most evident for model $A0$. As
shown in Table~\ref{table2} the integral over the right cut is
still finite but increased by a factor $M^2/m^2$ as compared to Table~\ref{table1}.
However, this factor is canceled by the finite contribution of the
circle at infinity. The latter contribution vanishes for all other
models. Table~\ref{table2} shows the increase of the
right-cut integral by $M^2/m^2$ also by
all the other models. The somewhat larger factor for $C0$ is due to the bad
convergence in this model. However, also the contributions of the
residues and the left cut have much increased, and as a result
the full contour integral yields precisely the value predicted by
the real part of the functions. Because of the additional factor $s$ with respect
to the forward polarizability, a figure analogous to Fig.~\ref{fig:modelC0}
is completely dominated by the spurious pole at $s \approx -11~{\text {GeV}}^2$
such that the effect of the $\rho$ meson is practically invisible. As a result the
residue of the spurious pole cancels the contribution of the right cut nearly completely. In a similar
way, the spurious left cut due to the $1/ \sqrt{s}$ ansatz in model
$F_C$ cancels 85~\% of the contribution of the (physical) right cut.
As required by Eq.~(\ref{eq:models}), we find the relation
$P^{(+)}(s=m^2)=-P^{(-)}(s=m^2)$ not only for the real parts of the
model amplitudes but also by summing up all the contributions to the contour
integral. In spite of relatively small differences among the models near the resonance region,
the polarizabilities predicted  by considering only the right cut differ
by an order of magnitude. We conclude that the extrapolation from the resonance
position to the threshold of Compton scattering is dangerous, in particular, if
performed with functions containing unphysical singularities.
Because the $\omega$ and $\sigma$ mesons are the most important
agents in the dispersion analysis of
Refs.~\cite{Fil'kov:2005ss,Fil'kov:2005yf}, we have also studied
these cases. The numerical calculations for the $\omega$ meson are
performed for $M=M_{\omega}=0.782$~GeV, $m=M_{\pi^0}=0.135$~GeV,
$\Gamma_0=\Gamma_{\omega}=8.41$~MeV, and $\Gamma_{\gamma
\pi}=\Gamma_{\omega \gamma \pi}=0.715$~MeV. Except for its larger
photon decay branch and the smaller width, the $\omega$ follows the
above exercise for the $\rho$ very closely, see Tables~\ref{table3}
and \ref{table4}. In particular we observe the same overestimation
of $\alpha-\beta$ by the dispersion integral along the right cut in the
$s$ channel.\\

The scalar $\sigma$ meson is the dominant agent for the $t$-channel
reaction. In the following we describe this meson with the parameters of
model (a) given by Table 1 of Ref.~\cite{Fil'kov:1998np}, in particular
$M=M_{\sigma}=0.547$~GeV, $m=M_{\pi^+}=0.140$~GeV,
$\Gamma_0=\Gamma_{\sigma}=1.204$~GeV, and $\Gamma_{\gamma
\pi}=\Gamma_{\sigma \gamma \pi}=0.62$~keV. Our models $A0$, $B$,
and $C0$ are defined as above, that is with an energy-independent
coupling constant fixed at resonance and with propagators given by
Eqs.~(\ref{eq:FA}) - (\ref{eq:Gamma-s}). The models $A$, $B$, and $C$
contain an energy-dependent coupling constant and for models $B$ and
$C$ an additional energy-dependence of the width involving ${\sqrt t}$
factors. In particular, model $C$ corresponds to the expressions found
in the Appendix of Ref.~\cite{Fil'kov:1998np}. Because of the spin
$J=0$ of the $\sigma$, the width opens like a square root at
threshold. The low mass and the huge width of the $\sigma$ lead to
quite different analytical structures as compared to the vector
mesons. Whereas models $A$ and $A_0$ with their constant widths
yield a pole at the complex mass $t=(M-i\, \Gamma_0/2)^2$, this pole
has been moved onto the second sheet for $B$ and $B0$, and in the
case of $C$ and $C0$ the $\sigma$ pole has completely disappeared.
The assumed analytical form also leads to a reasonable
convergence of the dispersion integral over the right cut. However,
the ansatz of Ref.~\cite{Fil'kov:1998np} results in a left cut
with a singularity at $t=0$ leading to a dispersion integral with an
integrand like $t^{-3/2}$ near the origin. As a consequence the
integral diverges like $t^{-1/2}$ for $t \rightarrow 0$, at which
point the polarizability is defined. The numerical results are
listed in Table~\ref{table5}. \\
\section{Dispersion relations in the $t$ channel}
\label{sec:t-channel}
The amplitudes $A_1$ and $A_2$ introduced in Sec.~\ref{sec:kinematics} correspond to
photon helicity differences $\Lambda_\gamma=0$ and $\Lambda_\gamma=2$, respectively.
They have the following partial-wave expansions involving total angular momentum
$J\ge \Lambda_\gamma$ and even isospin $I=0,\,2 $:
\begin{eqnarray}
A_1^I&=&\sum_{J\ge0} \sqrt{2J+1} {\cal A}_{J0}^I P_J^0(\cos\theta^*),
\nonumber\\
\quad A_2^I&=&\sum_{J\ge2} \sqrt{2J+1}\sqrt{\frac{(J-2)!}{(J+2)!}}
{\cal A}_{J2}^I \frac{P_J^2(\cos\theta^*)}{(1-\cos^2\theta^*)},
\end{eqnarray}
where $\theta^*$ is the scattering angle in the c.m.~system. The isospin decomposition
of the physical channels is given by
\begin{eqnarray}
A_i(\gamma\gamma\rightarrow\pi^+\pi^-)=\sqrt{\frac{1}{3}}A_i^{I=0}
+\sqrt{\frac{1}{6}}A_i^{I=2},\nonumber\\
A_i(\gamma\gamma\rightarrow\pi^0\pi^0)=\sqrt{\frac{1}{3}}A_i^{I=0} -\sqrt{\frac{2}{3}}A_i^{I=2}.
\end{eqnarray}
The partial waves ${\cal A}_{J\Lambda_\gamma}^I$ correspond to eigenstates of the scattering matrix,
and their imaginary parts can be constructed by unitarity as follows:
\begin{equation}
{\rm Im}{\cal A}_{J\Lambda_\gamma}^I(\gamma\gamma\rightarrow\pi\pi)= \sum_n\rho_n
{\cal A}_{J\Lambda_\gamma}^I(\gamma\gamma\rightarrow n) {\cal I}_{J\Lambda_\gamma}^I
(n\rightarrow \pi\pi), \label{eq:unitarity}
\end{equation}
with $\rho_n$ the density of states for each channel $n$ and ${\cal I}$ the
hadronic amplitude for the decay $n\rightarrow \pi\pi$. In the elastic region, the sum on the right hand
side of Eq.~(\ref{eq:unitarity}) is saturated by the two-pion channel and, as an immediate consequence,
the phase $\phi^I_{J\Lambda_\gamma}$ of each partial wave equals the corresponding
pion-pion phase shift $\delta^I_J\equiv{\rm arg}\,{\cal I}^I_J(\pi\pi\rightarrow
 \pi\pi)$. This fact can be incorporated into the Omn\`{e}s function, which is
constructed to have the phase of the $\pi\pi$ scattering amplitude above two-pion
threshold and to be real otherwise,
\begin{equation}
{\Omega^I_J}(t)\;=\;\exp{\left[ \frac{t}{\pi}\;{\int_{4m^2}^{\infty}}\,dt'\,
\frac{{\delta^{IJ}_{\pi\pi}}(t')}{t'(t'\,-\,t\,-\,i\varepsilon)}\right]} \;.
\end{equation}
The function ${\cal A}^I_{J \Lambda_\gamma}(t)/\Omega^I_J(t)$ is by construction real above
$\pi\pi$ threshold and has a left cut ${\cal H}^I_{J\Lambda_{\gamma}}(t)$
generated for $t < 0$ by the Born term and for $t < -(M_V^2-m^2)/M_V^2$ by the exchange
of vector mesons like the $\rho$ and $\omega$ vector mesons in the $s$ and $u$ channels.
Hence the difference $\left[{\cal A}_{J\Lambda_\gamma}^I(t) -{\cal H}^I_{J\Lambda_\gamma}(t)\right]
/\Omega^I_J(t)$ has only a right cut from $t=4m^2$ to $\infty$, and
satisfies the following dispersion relation:
\begin{eqnarray}
{\cal A}^I_{J \Lambda_\gamma}(t)\,&=&\,\Omega^I_J(t)\;\bigg\{ {\cal H}^I_{J \Lambda_\gamma}(t)\,
\mathrm{Re} \left[(\Omega^I_J)^{-1}(t)\right] \nonumber\\
&-&\frac{(t\,-\,4m^2)^{(J-\Lambda_\gamma)/2}}{\pi}\, \int^{\infty}_{4m^2} \,dt'\,
\frac{{\cal H}^I_{J \Lambda_\gamma}(t')\, \mathrm{Im}\left[(\Omega^I_J)^{-1}(t')\right]}
{(t'\,-\,4m^2)^{(J-\Lambda_\gamma)/2}(t'\,-\,t)}
\bigg\}\,,
\label{eq:gagapipidisprel}
\end{eqnarray}
with the factor $(t'-4m^2)^{(J-\Lambda_\gamma)/2}$ providing the right asymptotic behavior for the
convergence of the integral. In particular, the S-wave amplitude is given by
\begin{equation}
{\cal A}^I_{00}=\Omega^I_0(t)\,\bigg\{ {\cal H}^I_{00}(t)\, \mathrm{Re} \left[(\Omega^I_0)^{-1}(t)\right]
-\frac{1}{\pi}\, \int^{\infty}_{4m^2} \,dt'\, \frac{{\cal H}^I_{00}(t')\, \mathrm{Im}\left[(\Omega^I_0)^{-1}(t')\right]}
{(t'\,-\,t)}\bigg\}. \label{eq:swave_uns}
\end{equation}
Furthermore, we define the generalized Born term as
\begin{equation}\label{eq:generalBorn}
H_i=A_i^B+A^V_i,
\end{equation}
with $A_i^B$ the Born term of Eq.~(\ref{eq:Born}) and the vector-meson contributions
$A^V_i$ given by Ref.~\cite{Ko:1989yd}
\begin{eqnarray}
A^V_1(s,t)&=&-2e^2R_V\left[\frac{s}{s-M_V^2}+\frac{u}{u-M_V^2}\right],\label{eq:A1V}\\
A^V_2(s,t)&=&2e^2m^2 R_V\left[\frac{1}{s-M_V^2}+\frac{1}{u-M_V^2}\right],
\label{eq:A2V}
\end{eqnarray}
where $R_V$ are determined from the condition~\cite{Donoghue:1993kw}
\begin{equation}
R_V= \frac{24 \pi M^3_V}{e^2}\frac{\Gamma(V\rightarrow\pi\gamma)}{(M_V^2-m^2)^3}\;.
\end{equation}
The comparison with Eqs.~(\ref{eq:FA}) - (\ref{eq:models}) in the zero-width
approximation yields the relation $e^2 R_V=4g^2_V(M_V^2)$, with $g_V^2(M_V^2)$ the
coupling of Eq.~(\ref{eq:g2-s}) at resonance. In the following we only discuss the most important
vector mesons, the $\rho(770)$ and $\omega(782)$.\\

At energies $W_{\pi \pi} = \sqrt{t} <1~{\text {GeV}}$ the phases are only large for the partial waves with $I=J=0$,
and therefore most of the final-state interaction is contained in $\Omega^I_{00}$. We construct
these S waves with the phase shifts given by Ref.~\cite{Froggat:1977dd}. The S-wave projections of the
Born and vector meson contributions read
\begin{eqnarray}
{\cal B}^{I=0}_{00}&=& \frac{1}{\sqrt{2}} {\cal B}^{I=2}_{00}= -2e^2\frac{2}{\sqrt{3}}\frac{1-\beta^2}{t\beta}
{\rm ln}\left(\frac{1+\beta}{1-\beta}\right),\\
{\cal V}^{I=0}_{\rho\,00}&=&4 e^2 \sqrt{3}\frac{R_\rho}{t}
\left[\frac {M_{\rho}^2}{\beta} {\rm ln}\left(\frac{1+\beta+t_\rho/t}{1-\beta+t_\rho/t}\right)-t\right],\\
{\cal V}^{I=2}_{\rho\,00}&=&0,\\
{\cal V}^{I=0}_{\omega\,00}&=& -\frac{1}{\sqrt{2}}{\cal V}^{I=2}_{\omega\,00}= \frac{4 e^2}{\sqrt{3}}\frac{R_\omega}{t}
\left[\frac {M_{\omega}^2}{\beta}  {\rm ln}\left(\frac{1+\beta+t_\omega/t}{1-\beta+t_\omega/t}\right)-t\right],
\end{eqnarray}
where $t_V=2(M^2_V-m^2)$ and $\beta=\sqrt{1-4m^2/t}$.\\

Because the phases of the higher partial waves are generally small for $W_{\pi \pi}<1~{\text {GeV}}$, they can be
well described by the generalized Born amplitude in this region, and their contribution
to the polarizability can be neglected. However, the $\gamma\gamma\rightarrow \pi\pi$ process
has a distinct resonance structure corresponding to the isoscalar $f_2(1270)$ resonance with
mass $M_{f_2}=1275$~MeV and width $\Gamma_{f_2}=185$~MeV.  This resonance shows up in the partial wave
${\cal A}_{22}^0$ and therefore contributes only to the amplitude $A_2^0$. We model this contribution
according to Ref.~\cite{Drechsel:1999rf}
by the Breit-Wigner ansatz
\begin{equation}
A_2^{f_2} \;=\;e^2\, \frac{m^2}{M_{f_2}^2}\, \frac{g_{f_2 \gamma\gamma} \,
g_{f_2 \pi\pi}}{M_{f_2}^2\,-t\,-\,iM_{f_2} \,\Gamma_0}\;,
\label{eq:f2}
\end{equation}
where the coupling constant $g_{f_2\pi\pi}$ is known from the decay $f_2\rightarrow \pi\pi$,
$g_{f_2\pi\pi}=23.65$~\cite{Yao:2006px}, and $g_{f_2\gamma\gamma}$ is fitted to the
$\gamma\gamma\rightarrow \pi\pi$ cross section at the resonance position, resulting in
$g_{f_2\gamma\gamma}\approx 0.247$ consistent with Ref.~\cite{Yao:2006px}. As a result the $f_2$
resonance contribution is $ (\alpha+\beta)^{\pi^0}_{f_2}=
(\alpha+\beta)^{\pi^+}_{f_2}=0.17$.\\

\begin{figure}[ht]
\begin{center}
\epsfig{file=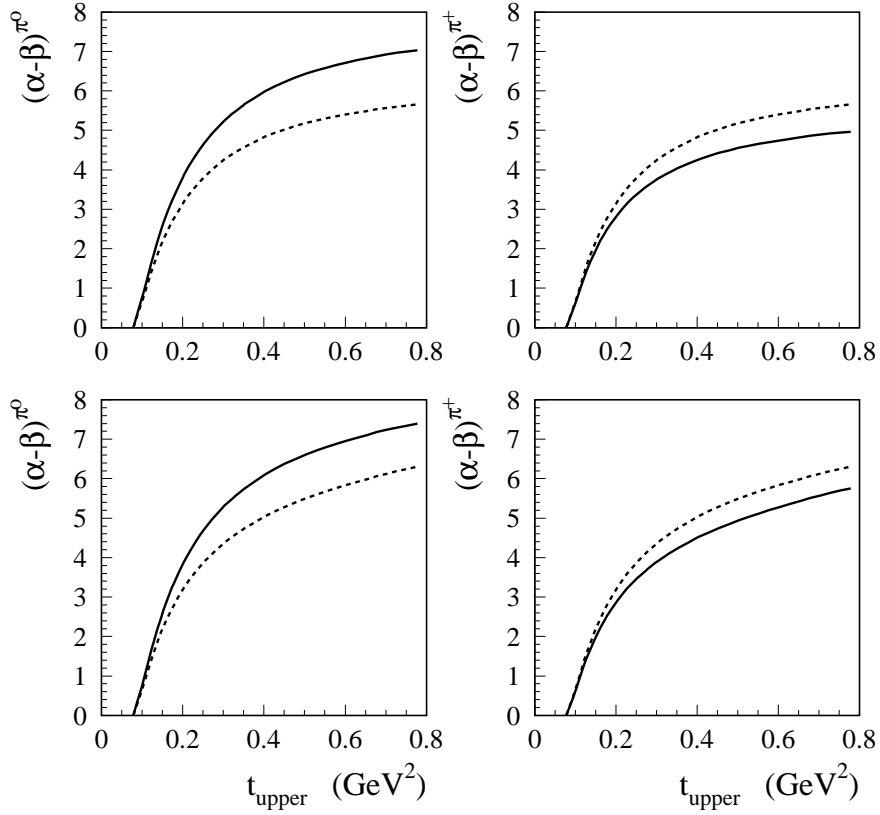, width=12 cm}
\end{center}
\caption{The $t$-channel two-pion contribution to the backward polarizability $\alpha - \beta$
of the $\pi^0$ (left panels) and $\pi^+$ (right panels) as a function of the upper limit of
integration $t_{\text {upper}}$. Top row: results obtained from the unitarized Born contribution,
bottom row: results obtained from the unitarized generalized Born contribution including also
the $\rho$ and $\omega$ contributions. Dashed lines: contribution of the $I=0$ channel,
solid lines: total result for the sum of the $I=0$ and $I=2$ contributions. All polarizabilities in
units of $10^{-4}~{\text {fm}}^3$.}
\label{fig:integral(alpha-beta)}
\end{figure}

With the unitarized S-wave contribution from Eq.~(\ref{eq:swave_uns}) and,
for the higher partial waves, the generalized Born contribution of Eq.~(\ref{eq:generalBorn}) and
the $f_2$ resonance contribution of Eq.~(\ref{eq:f2}), the full $t$-channel amplitudes
can be cast into the form
\begin{eqnarray}
A^I_{1}(s,t)\,&=&{\cal A}^I_{00}-{\cal H}^I_{00}(t)+H^I_1(s,t),\nonumber\\
A^I_{2}(s,t)\,&=&\, H^I_2(s,t)+A_2^{I \, f_2}(s,t). \label{eq:fullampl}
\end{eqnarray}
These amplitudes lead to the following results for the polarizabilities:
\begin{eqnarray}
\alpha-\beta&=&-\frac{1}{4\pi m} \left(A_1^V(m^2,0) -\frac{1}{\pi}\int^{\infty}_{4m^2} \,dt'\,
\frac{{\cal H}^I_{00}(t')\mathrm{Im}\left[(\Omega^I_0)^{-1}(t')\right]}{t'}\right),\label{eq:alpha_min_beta}\\
\alpha+\beta&=&-\frac{1}{4\pi m} (A_2^V(m^2,0)+A_2^{f_2}(m^2,0)).\label{eq:alpha_plus_beta}
\end{eqnarray}
Because Eqs.~(\ref{eq:A1V}) and (\ref{eq:A2V}) yield $A_1^V(m^2,0)=-A_2^V(m^2,0)>0$, it follows from
the above equations that the vector mesons contribute only to the magnetic polarizability $\beta$.
This effect is paramagnetic because the quark spins are aligned in the transition $\pi \rightarrow V$.
If we compare with the results of Sec.~\ref{sec:models}, the vector-meson contributions in
Eqs.~(\ref{eq:alpha_min_beta}) and (\ref{eq:alpha_plus_beta}) correspond to the $s$-channel term
in the fixed-$u$ dispersion relation of Eq.~(\ref{eq:DRu}), whereas the dispersion integral in
Eq.~(\ref{eq:alpha_min_beta}) and the $f_2$ resonance contribution in  Eq.~(\ref{eq:alpha_plus_beta})
can be identified with the $t$-channel dispersive contribution in Eq.~(\ref{eq:DRu}).
Focussing on the $t$-channel term, Fig.~\ref{fig:integral(alpha-beta)} shows the difference of
the polarizabilities $\alpha - \beta$ of Eq.~(\ref{eq:alpha_min_beta}) as a function of
the upper limit of integration in the region $4m^2 \leq t_{\text {upper}} \leq 0.78~{\text {GeV}}^2$.
The latter value is defined by the onset of inelasticities in the $\pi\pi$ phase shifts. As shown by
the figure, the $I=0$ channel (with the quantum numbers of the $\sigma$ meson) provides a contribution
of only about 5 units to the backward polarizability, about half of the value predicted by~\cite{Fil'kov:2005ss}.
The bottom panels show the results for the generalized Born term
${\cal H}^I_{00}={\cal B}^I_{00}+{\cal V}^I_{00}$, including the $\rho$ and $\omega$ contributions,
whereas the top panels are obtained with only the Born term ${\cal B}^I_{00}$. Furthermore,
Fig.~\ref{fig:integral(alpha-beta)} shows the results for both the $I=0$ channel only (dashed lines)
and the sum of the $I=0$ and $I=2$ channels (solid lines).\\

\begin{figure}[ht]
\begin{center}
\epsfig{file=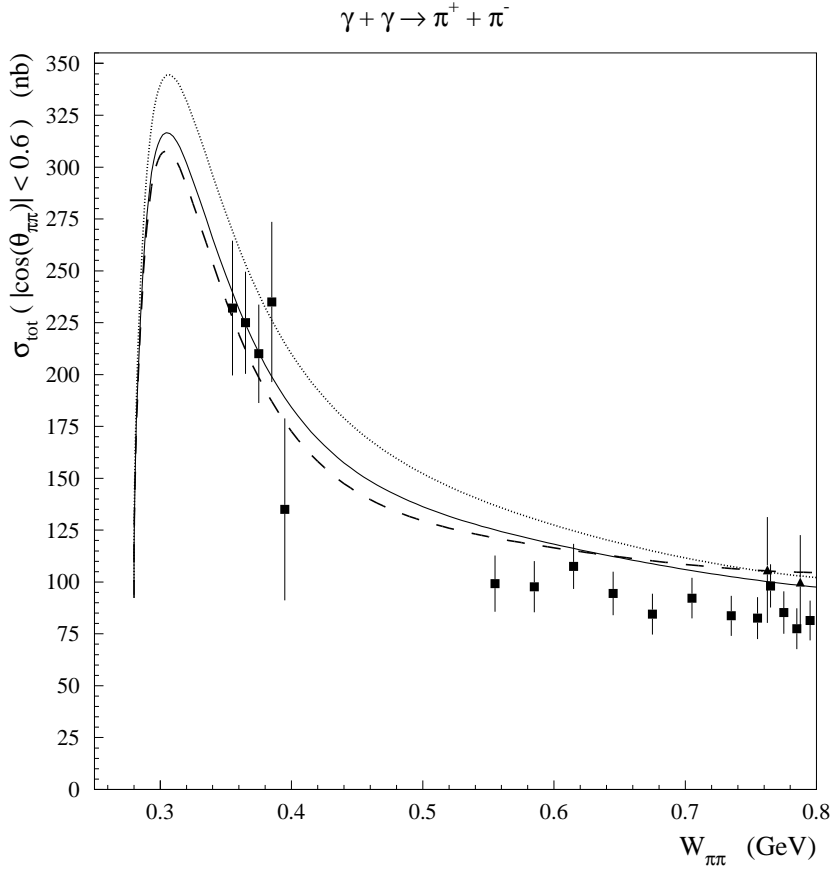,width=12 cm}
\end{center}
\caption{The total cross section for $\gamma\,\gamma \rightarrow \pi^+\,\pi^-$ as a function
of the c.m.~energy $W_{\pi \pi}=\sqrt{t}$ in the low-energy region, as obtained from the
unitarized generalized Born term (including the $\rho$ and $\omega$ contributions). Solid line: unsubtracted DRs,
dashed line: subtracted DRs with the subtraction constants given by the polarizabilities predicted
by the two-loop calculation of ChPT, dotted line: subtracted DRs with the subtraction constants
given by the polarizabilities obtained from unsubtracted DRs by Ref.~\cite{Fil'kov:2005yf}, except that the
vector-meson contribution is calculated with an energy-independent coupling constant $g_V(M_V^2)$.
The data are from J.~Boyer {\emph {et al.}}~\cite{Boyer:1990vu}.}
\label{fig:tot(pi+pi-)_vm}
\end{figure}

\begin{figure}[ht]
\begin{center}
\epsfig{file=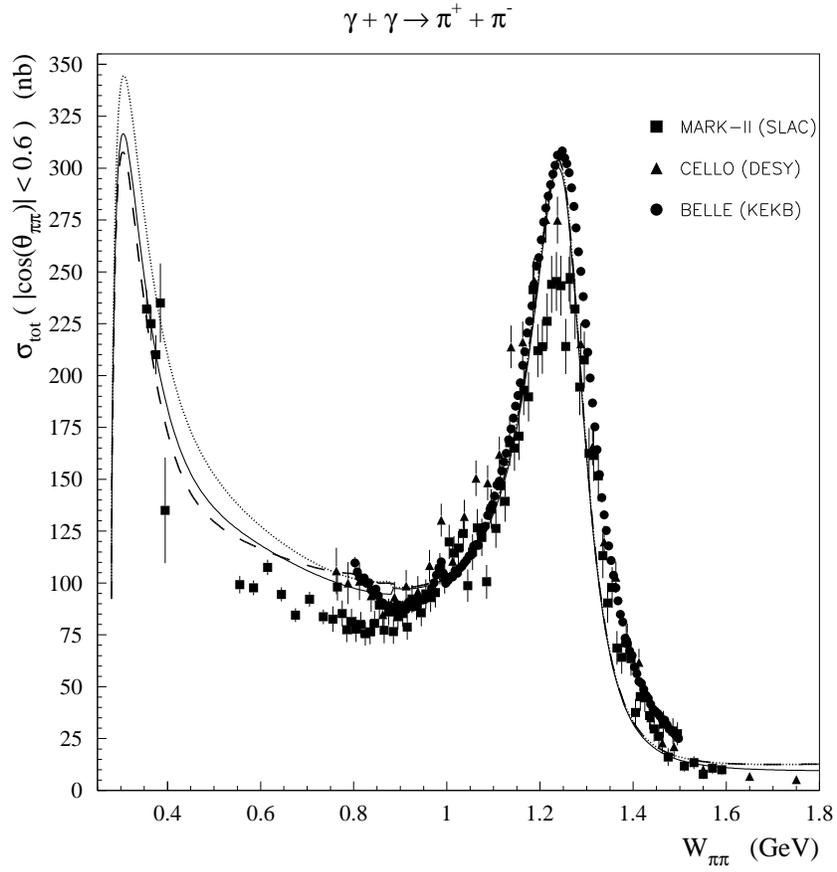, width=12 cm}
\end{center}
\caption{The total cross section for $\gamma\,\gamma \rightarrow \pi^+\,\pi^-$ as a function of the
c.m.~energy $W_{\pi \pi}=\sqrt{t}$ including the high-energy region dominated by the $f_2$ resonance.
Data from the collaborations MARK-II~\cite{Boyer:1990vu}, CELLO~\cite{Berend:1992dd}, and
BELLE~\cite{Mori:2007bu}. The error bars show only the statistical errors.
Further notation as in Fig.~\ref{fig:tot(pi+pi-)_vm}.}
\label{fig:tot(pi+pi-)_vmf2}
\end{figure}

The $\gamma\gamma\rightarrow \pi\pi$ cross section is obtained from
the amplitudes of Eq.~(\ref{eq:fullampl}) with the S-wave contribution
evaluated by the following subtracted DR:
\begin{eqnarray}
{\cal A}^I_{0 0}(t)\,&=&\,\Omega^I_0(t)\; \bigg\{ {\cal H}^I_{0 0}(t)\,\mathrm{Re}
\left[(\Omega^I_0)^{-1}(t)\right] \nonumber\\
&+&c^I_{00}  -\frac{t}{\pi}\, \int^{\infty}_{4m^2} \,dt'\,
\frac{{\cal H}^I_{00}(t')\, \mathrm{Im}\left[(\Omega^I_0)^{-1}(t')\right]} {t'(t'\,-\,t)}\bigg\},
\label{eq:swave_sub}
\end{eqnarray}
where the subtraction constants $c^I_{00}$ are related to the polarizabilities by
\begin{eqnarray}
c^{I=0}_{00}&=&-8\pi m\frac{1}{\sqrt{3}}\left[(\alpha-\beta)^{\pi^+}
+\frac{1}{2}(\alpha-\beta)^{\pi^0}\right]-A_1^{V,I=0}(m^2,0),\\
c^{I=2}_{00}&=&-4\pi m \sqrt{\frac{2}{3}} \left[(\alpha-\beta)^{\pi^+}-
(\alpha-\beta)^{\pi^0}\right]-A_1^{V,I=2}(m^2,0).
\end{eqnarray}
Figure~\ref{fig:tot(pi+pi-)_vm} displays the total $\gamma\gamma\rightarrow \pi^+\pi^-$
cross section given by the subtracted DRs of Eq.~(\ref{eq:swave_sub})
with subtraction constants fixed by ChPT and the model of Ref.~\cite{Fil'kov:2005yf}
as well as the results obtained from the unsubtracted DR for only the S-wave amplitude,
see Eq.~(\ref{eq:swave_uns}). We note that all the results are obtained with an
energy-independent coupling constant $g_V(M_V^2)$. The same results are shown over a larger energy
region in Fig.~\ref{fig:tot(pi+pi-)_vmf2}. The $f_2$ resonance contribution is clearly visible
near $W_{\pi \pi} =1.2$~GeV. However, the contribution of this resonance to the polarizability is very
small, as has been noted before.\\

\begin{figure}[ht]
\begin{center}
\epsfig{file=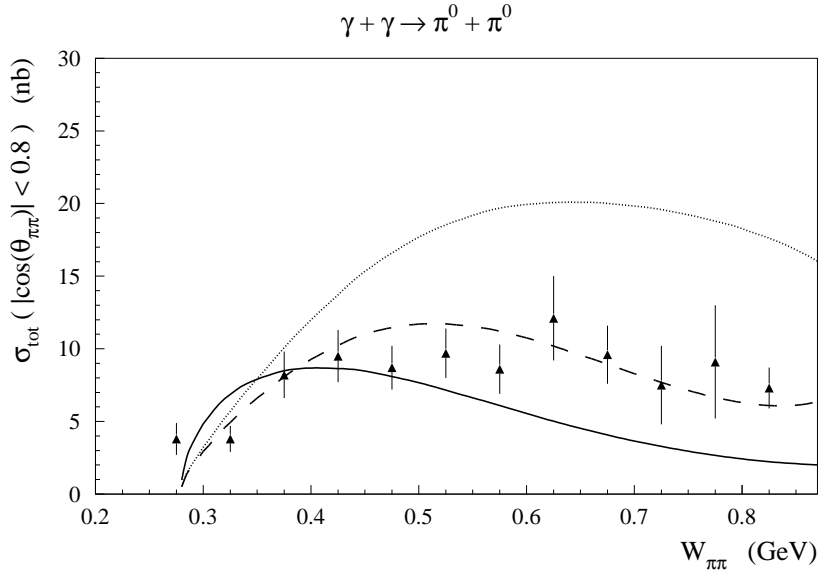,width=12 cm}
\end{center}
\caption{The total cross section for $\gamma\,\gamma \rightarrow \pi^0\,\pi^0$ as a function
of the c.m.~energy $W_{\pi \pi}=\sqrt{t}$ in the low-energy region. The data are from
H.~Marsiske {\emph {et al.}}~\cite{Marsiske:1990hx}.
Further notation as in Fig.~\ref{fig:tot(pi+pi-)_vm}.} \label{fig:tot(pi0pi0)_vm}
\end{figure}

\begin{figure}[ht]
\begin{center}
\epsfig{file=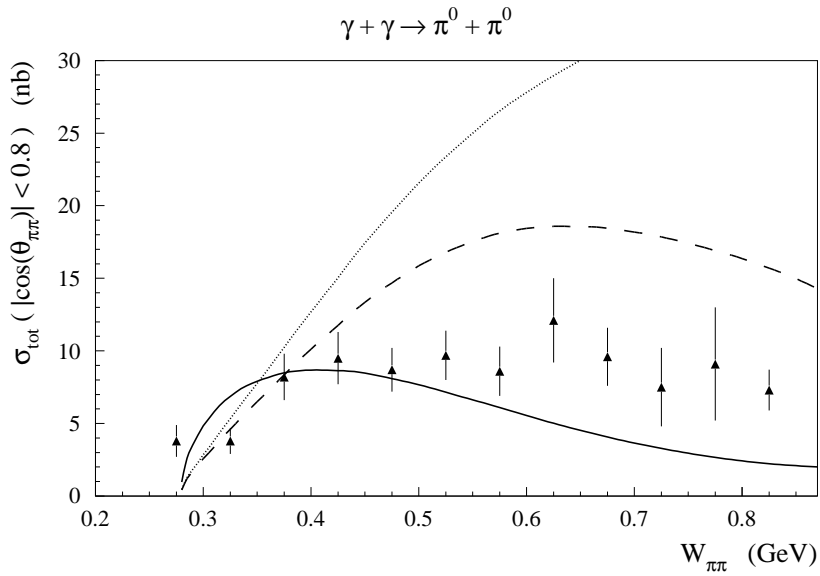, width=12 cm, height=8 cm}
\end{center}
\caption{The total cross section for $\gamma\,\gamma \rightarrow \pi^0\,\pi^0$ as a function
of the c.m.~energy $W_{\pi \pi}=\sqrt{t}$ in the low-energy region, as in Fig.~\ref{fig:tot(pi0pi0)_vm} but
obtained by unitarization of the Born terms, i.e., neglecting the vector meson contributions.
The data are from H.~Marsiske {\emph {et al.}}~\cite{Marsiske:1990hx}.
Further notation as in Fig.~\ref{fig:tot(pi+pi-)_vm}.}
\label{fig:tot(pi0pi0)}
\end{figure}

The corresponding results for the $\gamma\gamma\rightarrow \pi^0\pi^0$ cross section
are shown in Fig.~\ref{fig:tot(pi0pi0)_vm}. For this reaction
the differences among the models are much more pronounced, and at energies above the $f_2$ resonance
the discussed method fails completely, most likely because of the inelasticities due to more-pion
and heavier systems. In order to highlight the importance of the vector mesons,
Fig.~\ref{fig:tot(pi0pi0)} presents the results of the previous figure without
the vector-meson contributions. A correct unitarization of the full
amplitude will be required in order to describe the higher-energy region. Such a more
consistent treatment has been developed, for instance, in
Refs.~\cite{Kaloshin:1986yy} and \cite{Kaloshin:1993wj}. The large model dependency for the
neutral pion channel has also been observed in the recent work of Oller and Roca~\cite{Oller:2008kf}.
Finally, we present our predictions for $\alpha - \beta$ in Table~\ref{table6bis}, as obtained from
unsubtracted DRs. The value for the charged pion is in excellent
agreement with ChPT, whereas we fail to get close to the ChPT prediction for the neutral pion.
In view of the previous figures, this result is not too surprising. Whereas our unitarized
amplitude describes the $\pi^+$ cross section quite well up to energies of about 2~GeV
(see the solid lines in Figs.~\ref{fig:tot(pi+pi-)_vm} and \ref{fig:tot(pi+pi-)_vmf2}), the corresponding
results for the $\pi^0$ cross section are unsatisfactory (see the solid lines in
Figs.~\ref{fig:tot(pi0pi0)_vm} and \ref{fig:tot(pi0pi0)}). A comparison of the two latter
figures shows the large dependence on the background from meson exchange. Furthermore, the results
for the $\pi^0$ cross section from unsubtracted DRs do not describe the data for energies above 500~MeV.
A much better description of the data is obtained by the subtracted DR with the subtraction constant
as predicted by ChPT at the two-loop level~\cite{Bellucci:1994eb} (dashed line in Fig.~\ref{fig:tot(pi0pi0)_vm}).
\section{Conclusion}
\label{sec:conclusion}
The polarizabilities of the pion are elementary structure constants and therefore
fundamental benchmarks for our understanding of QCD in the confinement region.
Moreover, these polarizabilities have been calculated in ChPT to the two-loop order
with an estimated error of less than 20~\%. It is therefore disturbing that predictions
based on dispersion theory~\cite{Fil'kov:2005ss} yield $\alpha_{\pi^+} - \beta_{\pi^+}$
in the range of $11-15$, whereas ChPT~\cite{Gasser:2006qa} predicts $5.7 \pm 1.0$,
all in units of $10^{-4}~{\text {fm}}^3$. This discrepancy originates from
huge contributions of intermediate meson states in the approach of Ref.~\cite{Fil'kov:2005ss}. In particular,
the $\sigma$ exchange in the $t$ channel provides about 10 units for both neutral and charged
pions. For neutral pions, this $\sigma$ contribution is canceled by vector mesons, notably
$\omega$ exchange yielding a value of $-13$, with the result of
$\alpha_{\pi^0} - \beta_{\pi^0} \approx -3$. The large positive value for charged pions
results because (I) the $\omega$ does not contribute in this case and (II) axial vector mesons
provide additional positive contributions of about 4 units. In ChPT, on the other hand, the
vector mesons enter only at ${\cal {O}}(p^6)$ through vector-meson saturation of low-energy constants.
They are usually treated in the zero-width
approximation and estimated to yield a much smaller effect for the polarizability, e.g., the
$\omega$ contributes only about 1 unit to the neutral pion polarizability~\cite{Donoghue:1993kw,Bellucci:1994eb}.
The apparent discrepancy between the two approaches can be attributed to the specific forms
for the imaginary part of the Compton amplitudes~\cite{Fil'kov:2005ss,Fil'kov:2005yf}, which serve as input
for the dispersion integrals determining the polarizability at the Compton threshold
($s=m^2,\, t=0$). In order to quantify the strong model dependence of this procedure,
we have studied six different analytical forms including the model of Refs.~\cite{Fil'kov:2005ss,Fil'kov:2005yf}.
We recall that all these models are fitted to the same masses and widths of the exchanged mesons,
and in this sense they represent the experimentally known information in the same way. Concentrating
on the important $\omega$ meson and the forward polarizability, we find from Table~\ref{table3}
that the models $A0$, $B0$, and $C0$ (with a constant coupling strength) predict a
contribution of about 0.7 for the sum of $s$ and $u$ channel, in reasonable agreement with
Ref.~\cite{Donoghue:1993kw,Bellucci:1994eb}. The energy dependence of the coupling constant in
models $A$, $B$, and $C$ leads to an unphysical left cut (see Fig.~\ref{fig:modelC}) and
increases the contribution to the real part by a factor $M_{\omega} /m \approx 7$.
However, all the models agree if only the right cut is accounted for. If we now turn to the backward
polarizability, we expect from Eq.~(\ref{eq:models}) that the paramagnetic dipole transition involved
leads to a mere sign change compared to the forward polarizability, as is indeed reproduced by
the real part in Table~\ref{table4}. However, the right-cut integral has increased to large absolute values.
As a consequence, the dispersion integral over the right cut from two-pion threshold to infinity
does not converge to a (plausible) continuation of the real part to the
Compton threshold. The missing contributions to yield the real part are provided by unphysical
features of the models. We conclude that the strong singularities in the
form of poles (e.g., photon-pion ``bound states'') or unphysical cuts on the first Riemann sheet
are disturbing, because they violate the basic prerequisites of dispersion theory. For this reason
we do not see any conflict between dispersion theory {\emph {per se}} and ChPT. Of course, this does not exclude the
possibility of unexpected higher-order corrections in ChPT. However, our present knowledge on
vector mesons and their coupling to the electromagnetic
field does not indicate such large higher-order effects.\\

The arguments about the $\sigma$ exchange in the $t$ channel are more subtle. In Sec.~\ref{sec:models} we
have used the parameters of Ref.~\cite{Fil'kov:1998np} who put the pole position at
$M_{\sigma}=(0.547 - 0.602\,{\text{i}})$~GeV. The more recent analyses of pion-pion scattering find
such a resonance at $M_{\sigma}=(0.441 - 0.271\,{\text{i}})$~GeV~\cite{Caprini:2005zr} and
$M_{\sigma}=(0.456 - 0.241\,{\text{i}})$~GeV~\cite{Oller:2008kf}. Its large width of at least 500~MeV
and low mass (only about 300~MeV above two-pion threshold) lead to a complicated line-shape.
However, we consider it dangerous to model
this resonance with $1/\sqrt{t}$ factors~\cite{Fil'kov:1998np} because of the divergence
exactly at the point $t=0$ where the polarizability is determined. Instead we prefer the method outlined
in Sec.~\ref{sec:t-channel}, which follows Ref.~\cite{Donoghue:1993kw} and also previous calculations
using $t$-channel DRs to determine the nucleon's polarizability~\cite{Drechsel:1999rf}. In this way the
amplitudes are directly constructed from the pion-pion phase shifts, at least in the region below four-pion
threshold. Our numerical results yield an S-wave contribution of  about 6 units for both
charged and neutral pions, that is, only half of the value predicted for the
$\sigma$ contribution by Ref.~\cite{Fil'kov:1998np}.\\

Our calculations based on unsubtracted DRs for the Compton amplitudes are in reasonable agreement with ChPT for the charged
pion, whereas some agent is missing to reduce the backward polarizability of the neutral pion to the
small values predicted by ChPT. Contrary to the charged pion, the neutral-pion cross section is not
well described by unsubtracted DRs, and therefore no reliable prediction for $\alpha_{\pi^0} - \beta_{\pi^0}$ is
possible at present. The open questions in this field deserve further studies along the
lines presented in recent and ongoing work~\cite{Caprini:2005zr,Oller:2008kf}. At the same
time also new and independent experimental information as well as improved analysis of existing data are necessary.
We repeat that the polarizability of the pion is a fundamental benchmark of QCD in the realm of confinement.
It is therefore of utmost importance to clarify the yet existing discrepancies among the predictions of experiment
and theory.
\section*{Acknowledgments}
This research was
supported by the Deutsche Forschungsgemeinschaft (SFB443) and the EU Integrated
Infrastructure Initiative Hadron Physics Project under contract number
RII3-CT-2004-506078.


%

\begin{table*}
\begin{center}
\begin{tabular}{|c||c|c|c|c|}
\hline\hline
model & real part & right cut & left cut & residues \\
\hline
$A0$ & $0.036$ & $0.038$ & $-0.003$& $-$ \\
$B0$ & $0.037$ & $0.032$ & $-$     & $0.005$ \\
$C0$ & $0.037$ & $0.034$ & $-$     & $0.003$ \\
$A$   & $0.197$ & $0.045$ & $0.151$ & $-$ \\
$B$   & $0.205$ & $0.032$ & $0.170$ & $0.002$ \\
$C$   & $0.205$ & $0.033$ & $0.172$ & $-$ \\
\hline\hline
\end{tabular}
\end{center}
\caption{The $s$-channel contribution to the forward polarizability
$\alpha + \beta$ of the charged pion
predicted by different models for the $\rho$ meson propagator and
for the $\pi \rho \gamma$ coupling constant, see Eqs.~(\ref{eq:FA})
- (\ref{eq:g2-s}). The second column is
obtained directly from the real part of the model amplitudes, the
following columns list the different contributions from the
dispersion (contour) integral: the (physical) right cut, the
(unphysical) left cut, and the residues of the poles on the physical
sheet. Up to roundoff errors, the sum of the dispersive contributions (columns $\geq 3$)
adds up to the real part given in column 2. For model A0 describing a simple pole,
we have divided the integral over the real axis into the parts from $s=4m^2$ to $+\infty$
(listed as ``right cut'') and $-\infty$ to $s=4m^2$ (``left cut''). All numbers are in
units of $10^{-4}~{\text {fm}}^3$.} \label{table1}
\end{table*}
%
\begin{table*}
\begin{center}
\begin{tabular}{|c||c|c|c|c|c|}
\hline\hline
model & real part & right cut & left cut & residues & infinity\\
\hline
$A0$ & $-0.036$ & $-1.041$ & $-0.075$     & $-$     & $1.080$ \\
$B0$ & $-0.037$ & $-1.148$ & $-$     & $1.110$ & $-$ \\
$C0$ & $-0.037$ & $-1.928$ & $-$     & $1.891$ & $-$ \\
$A$   & $-0.197$ & $-1.062$ & $0.865$  & $-$ & $-$ \\
$B$   & $-0.205$ & $-1.020$ & $0.805$ & $0.010$ & $-$ \\
$C$   & $-0.205$ & $-1.177$ & $0.972$ & $-$ & $-$  \\
\hline\hline
\end{tabular}
\end{center}
\caption{The $s$-channel contribution to the  backward polarizability
$\alpha - \beta$ of the charged pion
predicted by different models for the $\rho$ meson propagator and
for the $\pi \rho \gamma$ coupling constant. The notation is the
same as in Table~\ref{table1}, except for the last column listing the contribution
of the big circle to close the contour at infinity.} \label{table2}
\end{table*}
%
\begin{table*}
\begin{center}
\begin{tabular}{|c||c|c|c|c|}
\hline\hline
model & real part & right cut & left cut & residues \\
\hline
$A0$ & $0.342$ & $0.343$ & $-0.001$& $-$ \\
$B0$ & $0.342$ & $0.342$ & $-$     & $-$ \\
$C0$ & $0.342$ & $0.342$ & $-$     & $-$ \\
$A$   & $1.980$ & $0.348$ & $1.631$ & $-$ \\
$B$   & $1.981$ & $0.340$ & $1.639$ & $-$ \\
$C$   & $1.981$ & $0.340$ & $1.639$ & $-$ \\
\hline\hline
\end{tabular}
\end{center}
\caption{The $s$-channel contribution to the  forward polarizability
$\alpha + \beta$ of the neutral pion
predicted by different models for the $\omega$ meson propagator and
for the $\pi \omega \gamma$ coupling constant. The notation is the
same as in Table~\ref{table1}.} \label{table3}
\end{table*}
%
\begin{table*}
\begin{center}
\begin{tabular}{|c||c|c|c|c|c|}
\hline\hline
model & real part & right cut & left cut & residues & infinity\\
\hline
$A0$ & $-0.342$ & $-11.426$  & $-0.042$    & $-$     & $11.126$ \\
$B0$ & $-0.342$ & $-12.389$  & $-$         & $12.047$& $-$ \\
$C0$ & $-0.342$ & $-22.584$  & $-$         & $22.245$& $-$ \\
$A$   & $-1.980$ & $-11.445$  & $9.465$     & $-$     & $-$ \\
$B$   & $-1.981$ & $-11.633$  & $9.039$     & $0.613$ & $-$ \\
$C$   & $-1.981$ & $-11.826$  & $9.846$     & $-$     & $-$  \\
\hline\hline
\end{tabular}
\end{center}
\caption{The $s$-channel contribution to the backward polarizability
$\alpha - \beta$ of the neutral pion
predicted by different models for the $\omega$ meson propagator and
for the $\pi \omega \gamma$ coupling constant. The notation is the
same as in Table~\ref{table1}, except for the last column listing the contribution
of the big circle to close the contour at infinity.} \label{table4}
\end{table*}
%
\begin{table*}
\begin{center}
\begin{tabular}{|c||c|c|c|}
\hline\hline
model & real part & right cut & left cut \\
\hline
$A0$ & $-0.878$ & $5.882$  & $-6.760$    \\
$B0$ & $ 7.410$ & $7.410$  & $-$         \\
$C0$ & $ 8.789$ & $8.789$  & $-$         \\
$A$   & $\infty$ & $6.307$  & $\infty$     \\
$B$   & $\infty$ & $8.548$  & $\infty$     \\
$C$   & $\infty$ & $9.357$  & $\infty$     \\
\hline\hline
\end{tabular}
\end{center}
\caption{The $t$-channel contribution to the  backward polarizability
$\alpha - \beta$ of the pion
predicted by different models for the $\sigma$ meson propagator and
for the $\pi \gamma \sigma$ coupling constant. The notation is the
same as in Table~\ref{table1}. However note that there are neither
contributions from residua on the first sheet nor from the contour
at infinity.} \label{table5}
\end{table*}
\begin{table*}
\begin{center}
\begin{tabular}{|c||c|c||c|c|c|c||c|}
\hline\hline
$\alpha-\beta$ &\multicolumn{2}{|c||}{Born}& \multicolumn{2}{|c|}{gen. Born}
& \multicolumn{2}{|c||}{vector mesons}&sum\\
\hline
        &  I=0  & I=2   & I=0  & I=2   & I=0 & I=2 & \\
\hline
$\pi^+$ &  5.65 & -0.69 & 6.30 & -0.54 &  -0.065   & 0    & 5.70\\
\hline
$\pi^0$ &  5.65 & 1.38 & 6.30 & 1.10 &  -0.30   & -0.47    & 6.62\\
\hline\hline
\end{tabular}
\end{center}
\caption{The backward polarizability $\alpha - \beta$ of the charged and
neutral pions in units of $10^{-4}~{\text {fm}}^3$. The results are obtained
by unitarization of the $t$-channel Born amplitude as well as generalized Born
amplitude and from the $s$- and $u$-channel contributions of vector mesons
in the narrow-width approximation. The contributions of the isospins $I=0$
and $I=2$ are given separately. The last column gives the sum of the vector-meson
and dispersive contributions, as obtained from the generalized Born amplitude.}
\label{table6bis}
\end{table*}
%
%
%
%
%
%
%
%
\end{document}